\documentclass[pdflatex,Packages/sn-mathphys-num]{Packages/sn-jnl}

\usepackage{graphicx}%
\usepackage{multirow}%
\usepackage{amsmath,amssymb,amsfonts}%
\usepackage{amsthm}%
\usepackage{Packages/mathrsfs}%
\usepackage[title]{appendix}%
\usepackage[table]{xcolor}%
\usepackage{textcomp}%
\usepackage{Packages/manyfoot}%
\usepackage{booktabs}%
\usepackage{Packages/algorithm}%
\usepackage{Packages/algorithmicx}%
\usepackage{Packages/algpseudocode}%
\usepackage{siunitx}
\usepackage[normalem]{ulem}
\usepackage[innercaption]{Packages/sidecap}
\usepackage{pdfpages} 
\usepackage{pgffor} 

\makeatletter
\AtBeginDocument{\let\LS@rot\@undefined}
\makeatother

\def\supplementfilename{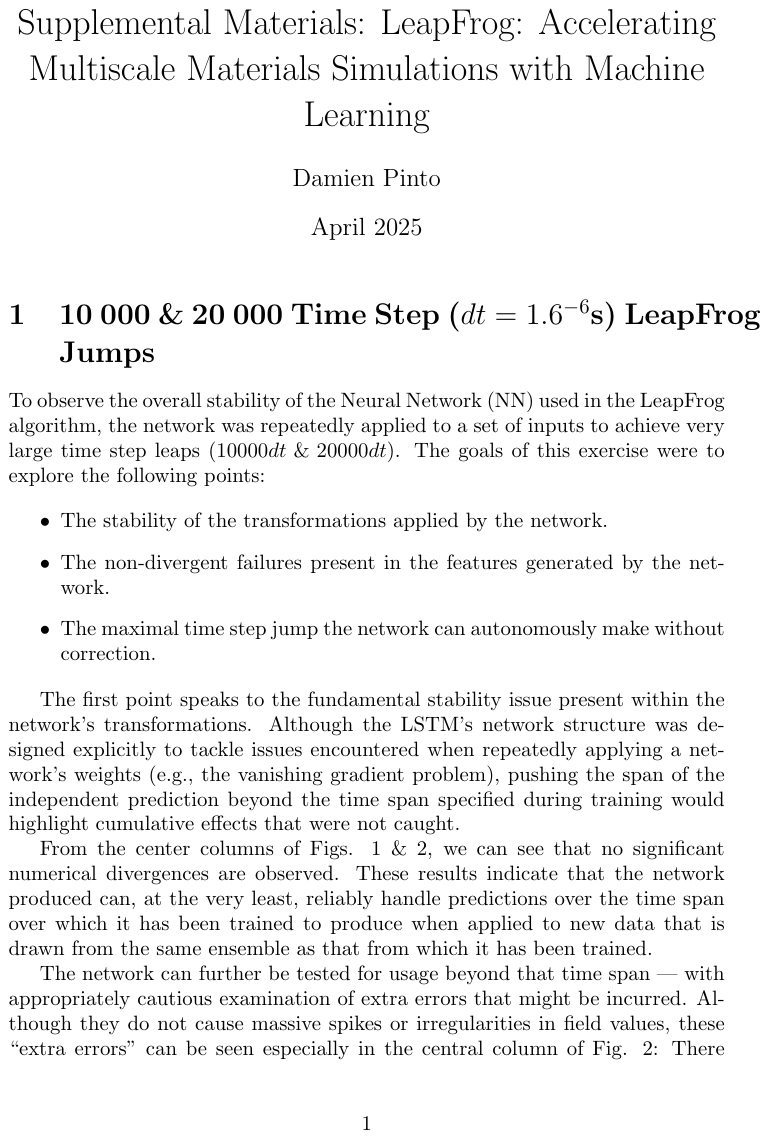}

\pdfximage{\supplementfilename}
\def\numbersupplementpages{\the\pdflastximagepages}

\newif\ifarXiv
\arXivtrue 

\begin{document}

\title[Article Title]{LeapFrog: Accelerating Multiscale Materials Simulations with Machine Learning}


\author*[1]{\fnm{Damien} \sur{Pinto}}\email{damien.pinto@mail.mcgill.ca}

\author[2]{\fnm{Michael} \sur{Greenwood}}\email{michael.greenwood@nrcan-rncan.gc.ca}

\author[1]{\fnm{Nikolas} \sur{Provatas}}\email{nikolaos.provatas@mcgill.ca}

\affil*[1]{\orgdiv{McGill Center for Materials Physics}, \orgname{McGill University}, \orgaddress{\street{3600 University}, \city{Montéal}, \postcode{H3A 2TB}, \state{Québec}, \country{Canada}}}

\affil[2]{\orgdiv{Natural Resources Canada}, \orgname{CanmetMATERIALs (CMAT)}, \orgaddress{\street{183 Longwood Road South}, \city{Hamilton}, \postcode{L8P 0A5}, \state{Ontario}, \country{Canada}}}

\abstract{Recent developments of novel materials have been greatly accelerated by computational modelling techniques that elucidate the complex physics controlling microstructure formation, the properties of which determine material function. The phase field (PF) method evolves said microstructure phases by coupling thermophysical and microscopic order parameter fields through multiple non-linear and costly to compute partial differential equations. Adaptive mesh refinement (AMR) significantly reduces the number of computations per time step, and thus the total computation time, by dynamically adapting numerical meshes resolution in proportion to local gradients. What AMR doesn't do is allow for adaptive time stepping. This work combines AMR with a neural network algorithm that uses a U-Net with a Convolutional Long-Short Term Memory (CLSTM) base to accelerate PF simulations. Our neural network algorithm is described in detail and tested on directional dilute binary alloy solidification simulations, a highly practical paradigm in alloy solidification. }

\keywords{Solidification, Multi Scale Modelling, Machine Learning, Phase Field Modelling, Materials}



\maketitle

\section{Introduction}\label{sec1}

The design of modern materials and their properties — be it mechanical, chemical, or electronic — relies on capturing the formation of its underlying microstructure during the solidification process. The formation of this microstructure is governed by the process parameters that describe the solidification process, e.g., temperature, pressure, concentration, cooling rate, etc.

Phase Field (PF) modelling, over recent decades, has become a linchpin of materials science thanks to its ability to capture the effects of these parameters over multiple length scales. This is achieved through the coupling of partial differential equations that describe the thermodynamic evolution of different fields within matter. For example, these fields can describe the concentration $C$ of substances within a sample, or even the state of matter present at different locations (the phase field $\phi$).

PF has been used to successfully tackle the formation of pure metals\cite{karmaRappelPF1996} and alloys\cite{echebarriaQttvAlloy2004}, whether in traditional casting or welding processes \cite{dantzigsolidification} or the rapid solidification rates found in additive manufacturing processes \cite{debroy2018additive}, as well as in solid state precipitation reactions \cite{zhu2004three}.

The aspects of PF that give it its strengths also comprise its most ubiquitous obstacle, however: a variety of length scales over many fields requires the manipulation of many high-resolution arrays at each time step of a simulation, and the size of said time steps is itself limited by the required spatial resolution. 

Luckily, theoretical and software advances have been able to take advantage of the recent boom in high-performance computing. The Adaptive Mesh Refinement (AMR) algorithm developed by Greenwood et al. \cite{greenwood_quantitative_2018} dramatically improves the allocation of computational resources by dynamically increasing the system resolution only where it is needed to accurately evolve the solidification interface. Other efforts have been made to leverage hardware developments by integrating Graphical Processing Units into the simulation pipeline, to significant effect\cite{yamanaka_gpu-accelerated_2011}\cite{guo_gpu-accelerated_2022}\cite{yang_gpu-accelerated_2017}\cite{kumar_ferrox_2023}.

Although these advancements have reduced the real-time costs of PF and expanded the system sizes and time scales it can access, the computational costs can nonetheless be prohibitive. AMR aided PF is still not immune from the limitations explicit methods impose on the time steps that can be taken. Implicit solvers can bypass this limitation at small system sizes, but not at the system sizes required for experimentally relevant simulations due to the number of nodes required.

As such, most experimentally relevant simulations still require weeks to perform on modern computational clusters. Moreover, \textit{multiple} simulations at one set of process parameters is needed to obtain an accurate microstructure characterization of systems simulated with stochastic noise. For this reason, there is still the need for further exploration and development of tools that can continue to push PF modelling forward.

Machine Learning is another tool that has gained attention, use, and development to great effect recently, and it has been demonstrated to be an apt complement to currently established computational tools \cite{yamanaka_gpu-accelerated_2011,greenwood_quantitative_2018,sakane2022parallel,wang2022application}. Specifically, in the field of materials design and solidification, Neural Networks (NN) have been successfully deployed in a generative capacity to simulations of amorphous carbon \cite{kilgour_generating_2020}, but also to accelerate the evolution of systems undergoing spinodal decomposition\cite{montes_de_oca_zapiain_accelerating_2021}\cite{hu_accelerating_2022}, 1st order phase transitions within small toy systems\cite{peivaste_machine-learning-based_2022}, the directional epitaxial growth of polycrystals\cite{qin_grainnn_2023}, and 3D grain evolution in additive manufacturing contexts\cite{choi_accelerating_2024}.

While there are numerous other examples predicting the time evolution of second order transformations (e.g., spinodal decomposition), and limited examples of a first order transformations with a single field theory, there are presently few that consistently couple both an order parameters (interface motion) with diffusion-limited transport — essential to better predicting accurate final microstructure length-scales and alloy concentration profiles?. 

There are significant differences, in terms of ML applications, between first and second order transformations. In 2nd order transformations, the universal scaling of space and time (e.g., domains growing according to $t^{\frac{1}{3}}$) present a clear pattern signal for a neural network to learn how spatial domains should evolve in time. No such universal scaling generally exists for first order transformations, and as a result there is a continuous need to maintain communication from the smallest scales (the interface width) to the higher length scales of thermal and solute diffusion that controls intercellular and side-branching spacing and morphology. Thus, exploration and development of ML practices that can address the multi-scale challenges of solidification problems are needed.

In this paper, we present a “LeapFrog” algorithm that combines AMR with a U-Net with a Convolutional Long-Short Term Memory (CLSTM) base trained with a novel dynamic multiscale loss metric that can accelerate the PF modelling of the important case of directional solidification of a dilute binary alloy with thermal noise and two-sided diffusion \cite{echebarriaQttvAlloy2004}. This algorithm is based on the original work of the thesis of the main author \cite{Pinto_MSC}. In Section \ref{sct:results} we present the wall time savings achieved by the algorithm, as well as the checks that establish the quantitative fidelity of the algorithm's output as well. In Section \ref{sct:discussion} we discuss very feasible further applications and expansions of the algorithm. Finally, in Section \ref{sct:methodology}, we outline the PF model used to generate the data, the design of the network's architecture, details of the multiscale loss, and the specifics of the LeapFrog algorithm.

\section{Results}\label{sct:results}
This section demonstrates the LeapFrog algorithm described in Section~\ref{methodology} and quantifies its time-acceleration features. We also review and quantify the fidelity with which the results of the LeapFrog algorithm match those of direct phase field simulations.

\subsection{Acceleration}

\begin{figure}
    \centering
    \includegraphics[width=0.85\textwidth]{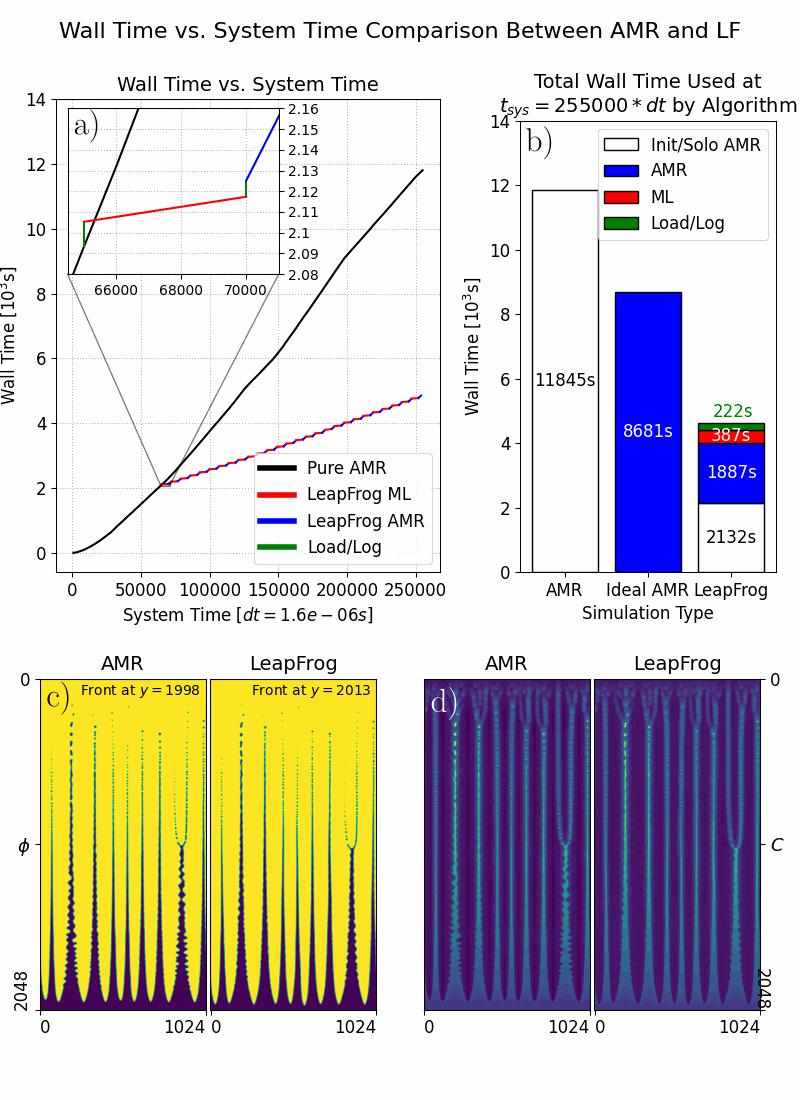}
    \caption{\textit{\textbf{5:2 Leapfrog Algorithm Results.} Comparison of the time savings and final results between traditional Adaptive Mesh Refinement (AMR) and neural network accelerated AMR phase field simulations, with $N_{ML}/N_{PF}=5/2$.\\\textbf{a)} System Time vs. Wall Time comparison between a Phase Field (PF) simulation using Adaptive Mesh Refinement (AMR) (black) and the LeapFrog algorithm (composite) evolution of the same system in the dendritic regime.\\\textbf{b)} Breakdown total Wall Time between PF+AMR, an “Idealized” scenario where the PF+AMR data is always organized in memory (minimizing lookup time), and the LeapFrog Algorithm (broken down by process).\\\textbf{c)} Comparison of the resulting $\phi$ and \textbf{d)} $C$ system fields after being evolved with the PF+AMR and the LeapFrog algorithm introduced in this work. 
    }}
    \label{fig:LFTime}
\end{figure}

A first example of the time savings of the LeapFrog algorithm are shown in Fig.~\ref{fig:LFTime}. The data show the results of an application of LeapFrog with $N_{ML} = 5 \cdot 1000dt$ and $N_{PF} = 2\cdot 1000dt$. It can be directly seen in panel a) that the overall wall time savings afforded by the LeapFrog algorithm increase the longer the system is evolved (in terms of system, i.e., physical, time). Even with the aforementioned overhead of the machine learning pipeline, the total evolution of the system, past some transient time, up to when the dendritic array reaches the end of the simulation domain, is achieved using less than half the wall time.

As noted in the latter portion of Section \ref{sct:MLArch}, this time comparison does not include the initial time needed to generate the training set and to train the initial network. During a parameter space exploration — the application for which the LeapFrog algorithm was designed for and where it could provide the largest benefit — \textit{multiple} simulation runs per process parameter set must be generated to obtain an ensemble view of the solidification microstructure. As a result, the first simulations generated can double for both microstructure characterization \textit{and} as a training set for the NN. Any subsequent simulation runs can then be accelerated using the trained network.

The pipeline just described is possible thanks to the network's accessible 7-hour training time on a training dataset composed of 11 simulation runs (in addition to 3 for the testing set, and 1 for the validation).

It is  noted that the efficiency of the LeapFrog algorithm is greater than an even “idealized” version of the Adaptive Mesh Refinement (AMR) algorithm (blue segments in panels a) \& b)). By “idealized” we mean a version of the AMR that retains its parallelized asynchronous execution of calculations while spatially adjacent cells in the system are \textit{always} assured to be adjacent, or at least close by, in memory.

In practice, results of computations are written to memory out of order in the standard AMR algorithm, which incurs an asymptotic increase in lookup time necessary to find a cell's neighbours for gradient computation. An optimal time interval could be determined where the overhead needed to reset the locations of cells in memory would result in an overall speed-up. However, for our purposes, this is not done here for the standard AMR algorithm.

LeapFrog always benefits from a quasi-ideal performance of the AMR mesh because machine learning predictions are always written to AMR input files in order. This is seen in frame a) of Fig.~\ref{fig:LFTime}, which shows that the slope of the AMR phase of the LeapFrog algorithm (data in blue) has a lower slope than the direct AMR algorithm, which as described above is not memory-cache optimized by default.
\begin{SCtable}
    \centering
    \begin{tabular}{|c|c|}
        \hline
        \rowcolor{white} \multicolumn{2}{|c|}{Average Wall Time per $1000dt$ of System Time} \\
        \hline
        \hline
        \rowcolor{white} \multicolumn{2}{|c|}{LeapFrog} \\
        \hline
        MSCLSTM & $2.24\si{\second}$\\
        \hline
        MSCLSTM (With Overhead) & $6.86\si{\second}$\\
        \hline
        \hline
        \rowcolor{white} \multicolumn{2}{|c|}{AMR} \\
        \hline
        AMR & $46.58\si{\second}$\\
        \hline
        Ideal AMR & $34.02\si{\second}$\\
        \hline
    \end{tabular}
    \caption{\textit{\textbf{LeapFrog Performance Metrics.} Comparison of wall times required per $1000dt$ of simulation time. When not indicated otherwise, the values listed are purely the time from the beginning of the computation of $1000dt$ to the end, not taking into consideration overhead (the time required to load up the system data and write the results to a file). The overhead required by the LeapFrog's current [proof-of-concept] implementation is the majority of its wall time cost.}}
    \label{tab:per1000Times}
\end{SCtable}

The overall microstructure evolution does present some differences between direct PF+AMR and LeapFrog-enabled PF algorithms. These mainly involve interface fluctuations affected by the presence of thermal noise. An example of this is seen in a dendrite extinction event between the last two dendrites on the right sides of frames c) \& d) of Fig.~\ref{fig:LFTime}. The formation of the halted (extinguished) tip appears slightly delayed between the two approaches, and the subsequent side-branch formation of the surviving dendrite arms is somewhat dampened in the LeapFrog algorithm. We expect that the discrepancy of the two approaches can be mitigated by balancing of $N_{ML}$ to $N_{PF}$ due to the absence of noise in the ML predictions. It is currently unclear how to incorporate to implement noise in the NN weights such as to, say, produce thermal fluctuations at interfaces, let alone satisfy the fluctuation-dissipation theorem.

\subsection{Adaptive Time Stepping}

\begin{figure}
    \centering
    \includegraphics[width=0.85\textwidth]{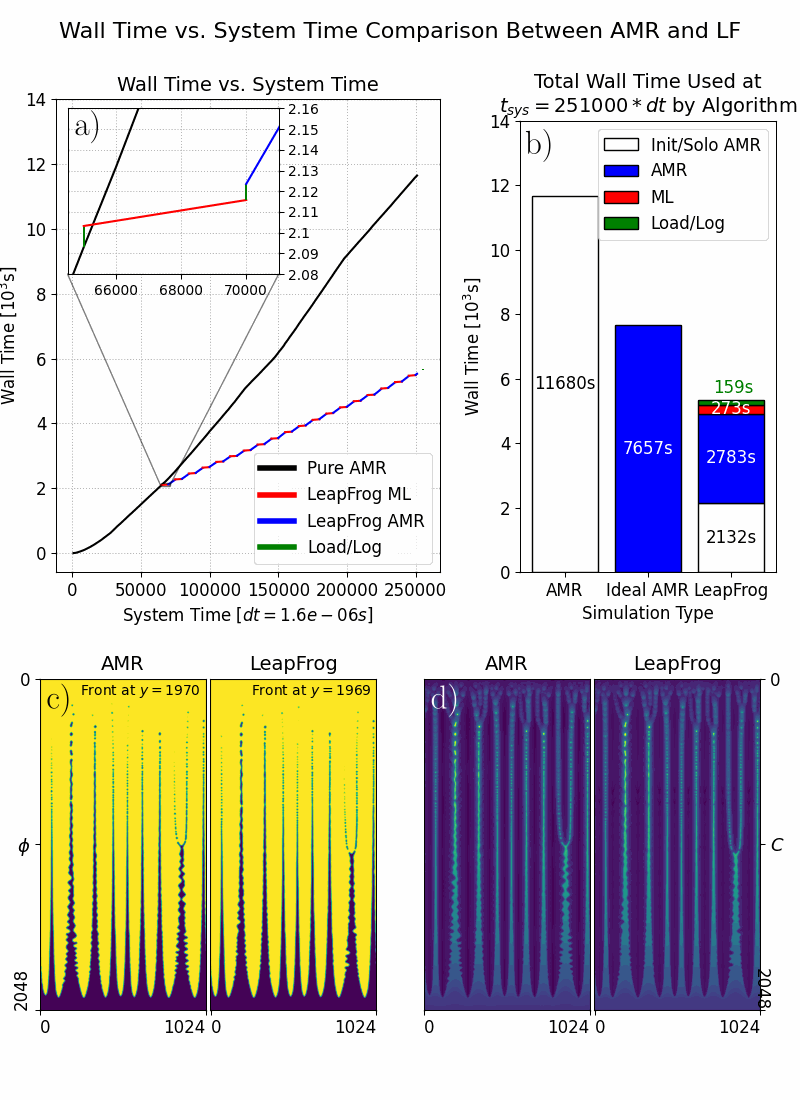}
      \caption{\textit{\textbf{5:5 LeapFrog Algorithm results.} Comparison of the time savings and final results between traditional Adaptive Mesh Refinement (AMR) and neural network accelerated AMR phase field simulations, with $N_{ML}/N_{PF}=1$.\\\textbf{a)} System Time vs. Wall Time comparison between a Phase Field (PF) simulation using Adaptive Mesh Refinement (AMR) (black) and the LeapFrog algorithm (composite) evolution of the same system in the dendritic regime.\\\textbf{b)} Breakdown total Wall Time between PF+AMR, an “Idealized” scenario where the PF+AMR data is always organized in memory (minimizing lookup time), and the LeapFrog Algorithm (broken down by process).\\\textbf{c)} Comparison of the resulting $\phi$ and \textbf{d)} system fields after being evolved with the PF+AMR and the LeapFrog algorithm introduced in this work. 
    }}
    \label{fig:LFTime5v5}
\end{figure}

To explore the adaptability of the LeapFrog algorithm in trading off between reducing simulation time and fidelity of produced solidification microstructure, another run is performed on the same initial seed as the one that produced the system in Fig. \ref{fig:LFTime}. This time, however, the algorithm is used with $N_{ML} = 5 \cdot 1000dt = N_{PF}$. The results are shown in Fig. \ref{fig:LFTime5v5}.

Going similarly through the panels, despite the less exploitative $N_{ML}/N_{PF}$ ratio, we can see in a) that we {\it continue} to obtain over a five-fold increase in simulation speed (per $1000dt$) leading to half the wall time needed to arrive at the end of the simulation domain. At the same time, we observe improved fidelity of side-branching and bulk/interface morphology well behind the solidification front, which demonstrates that the LeapFrog algorithm can be adjusted in its stepping ratio to adapt to the level of fidelity as required for the application. 

We have found that the LeapFrog-simulated microstructure for the data of Fig. \ref{fig:LFTime} remains true to the direct PF-simulated counterpart up to  $t=301000dt$, which illustrates the network adequately performing well past the temporal bounds of its training set ($t=250000dt$). This suggests that the transformations that the network has learned to emulate PF-solidification in a \textit{generalized} manner, not restricted to a location or time within the system's evolution.

Additional details on the tradeoff between simulation accuracy and acceleration provided by the LeapFrog algorithm — as well as an additional application of this kind of tuning of the algorithm — can be found in Supplementary Note 3.


\subsection{Prediction Quality of $\phi$ and $C$ Fields} \label{sct:predQlty}

While examinations of the overall morphology resulting from the evolution of a system through the LeapFrog algorithm is a good baseline benchmark, LeapFrog's utility lies also in its ability to provide \textit{quantitative} predictions. To that end, this Section examines the numerical fidelity of the resulting physical fields $\phi$ and $C$.

The first numerical characteristics we examined are profiles of the concentration field (segregation). The main interest is verifying the concentration profiles: (i) along the core (i.e., center) of one of the developed dendrites arms, (ii) along isotherms within the liquid, and (iii) along isotherms within the solid. The bulk concentration values within the core of the dendrites were used to verify that the LeapFrog algorithm produces output consistent with known physics such as the Gibbs Thomson and flux boundary conditions at play at a moving solid-liquid interface, which are satisfied to an excellent degree by the present Phase Field (PF) model \cite{echebarriaQttvAlloy2004}. The isotherm profiles, on the other hand, allow us to verify that the thermodynamics that underpin phase diagrams buried within the PF formulation is also respected.

\begin{figure}[t]
    \centering
    \includegraphics[width=0.95\textwidth]{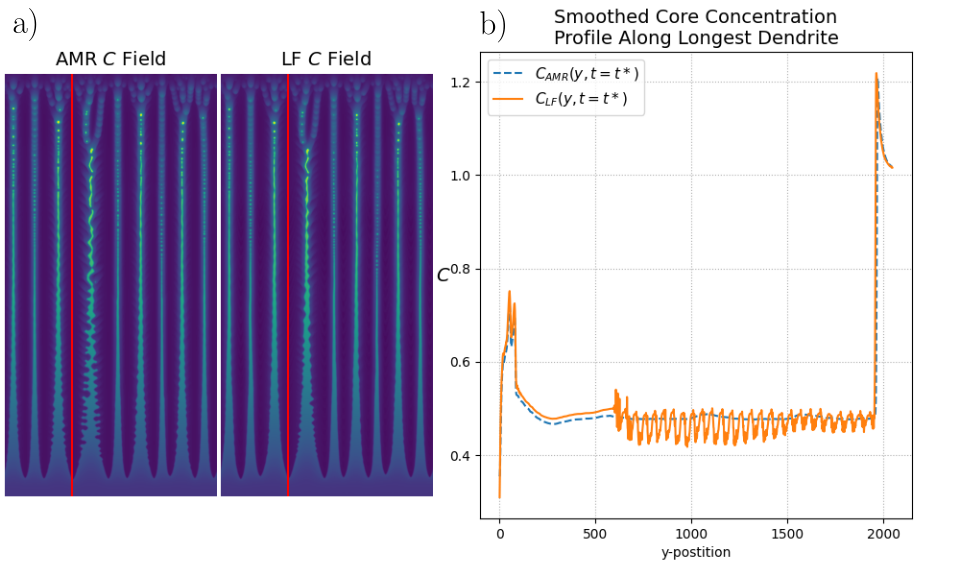}
    \caption{\textit{\textbf{PF and LeapFrog Concentration Profiles Comparison.} Comparison of concentration profiles along a dendrite's core (centerline) for two simulations, one evolved with Adaptive Mesh Refinement (AMR) (right frame, dashed line) and the other with the LeapFrog (LF) algorithm (right frame, continuous line). The left frame is split into two sub-frames, with the left sub-frame showing the centerline along which $C$ is measured from the AMR output, and the right sub-frame showing that of from the LeapFrog outputs. In the right frame. The point $y=0$ corresponds to the top of the centerline in the left sub-frames.} }
    \label{fig:CProfiles}
\end{figure}

\begin{figure}
    \centering
    \includegraphics[width=0.95\textwidth]{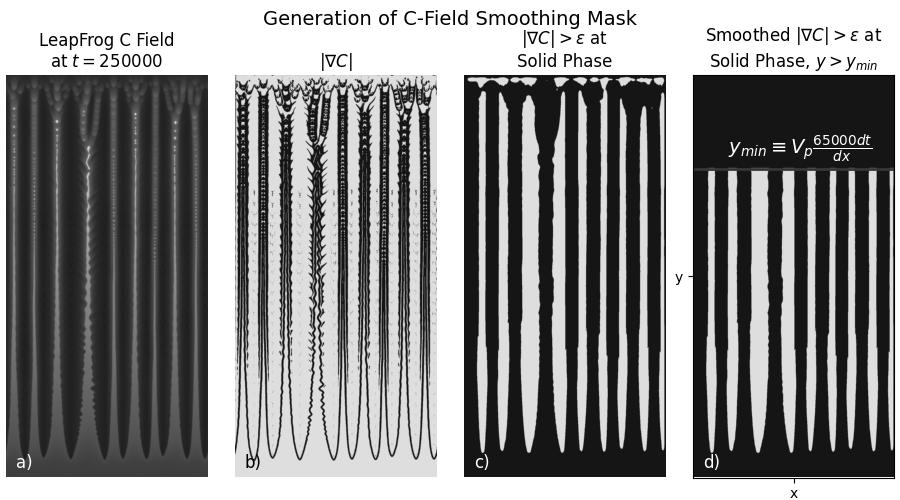}
    \caption{\textit{\textbf{Generation of mask used to smooth dendrite concentration profiles output from the LeapFrog algorithm}\\a) The concentration field of a system reaching the end of the dendritic growth regime.\\b) The magnitude of the gradients in said concentration field.\\c) Constraining the mask to areas where $\phi$ corresponds to the solid phase as well filtering the concentration gradient magnitude through a magnitude threshold.\\d) Restricting the resulting mask to the portion of the system generated after the initial transient phase of the simulation ($y > V_{p}\frac{65000dt}{dx}$).}}
    \label{fig:CSmoothMask}
\end{figure}

\begin{figure}[t]
    \centering
    \includegraphics[width=0.95\textwidth]{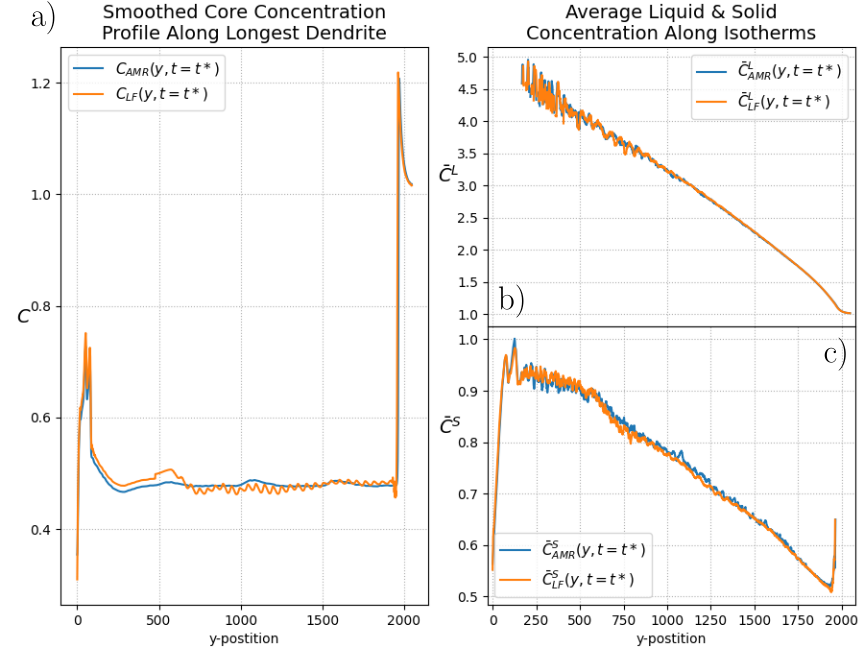}
    \caption{\textit{\textbf{Comparison of concentration profiles along a dendrite's core (centerline) for a simulation evolved with Adaptive Mesh Refinement (AMR) and LeapFrog (LF) at $t^{*}=250000dt$.} (red lines in Fig. \ref{fig:CProfiles}) with smoothing mask illustrated in Fig. \ref{fig:CSmoothMask} applied to the $C$-field output from the LeapFrog algorithm. The core (centerline) concentration profile is shown in a). Also shown are the concentration profiles along the system's isotherms in the liquid (b) and  solid (c) phases.}}
    \label{fig:CIsotherms}
\end{figure}

Initially, simply taking the outputs of the LeapFrog (LF) algorithm, we obtain the profiles illustrated in Fig. \ref{fig:CProfiles}. These exhibit small oscillations around the corresponding AMR solutions, as mentioned in Section \ref{sct:LossFctn}. 
We correct these oscillations by applying a smoothing filter to the area within the dendrites that are produced by the algorithm. This filter is implemented by producing a mask analogous to those used in feature weighted loss functions, illustrated in Fig. \ref{fig:CSmoothMask}, and applying a unitary smoothing operator at the indicated areas.
This outputs concentration profiles  almost exactly match the concentration profile predicted by direct PF simulations, as shown in Fig. \ref{fig:CIsotherms}. 


\begin{figure}[t]
    \centering
    \includegraphics[width=0.95\textwidth]{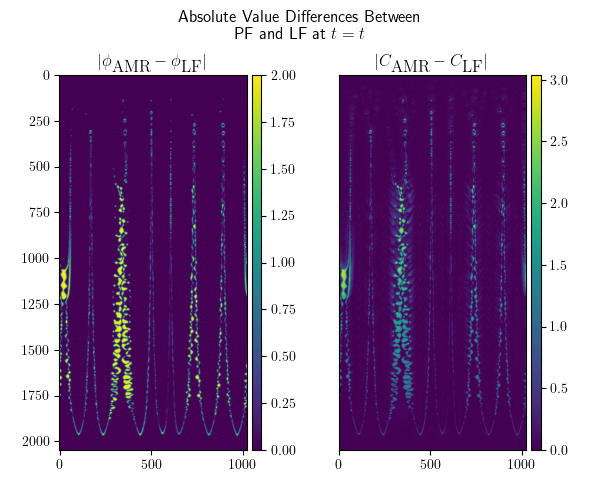}
    \caption{\textit{Absolute of pixel-wise error between the AMR and LeapFrog ($5000dt$ in ML, $2000dt$ PF alternation) runs at $t=250000dt$ for a run that shows particularly increased side-branching.}}
    \label{fig:absErr}
\end{figure}

\begin{figure}[t]
    \centering
    \includegraphics[width=0.95\textwidth]{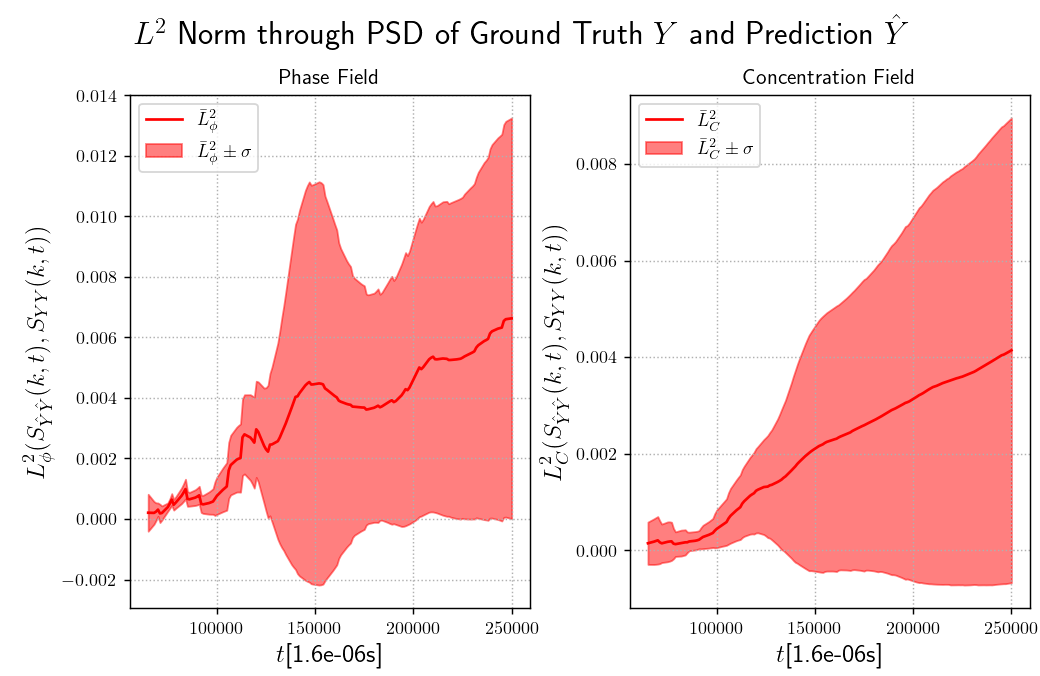}
    \caption{\textit{Mean and standard deviation of $L^2$ Norm distance between the Power Spectral Density(PSD)/power spectra of the interface as defined by the order parameter and concentration fields, averaged over 10 different AMR and LeapFrog runs.}}
    \label{fig:L2}
\end{figure}

We also examine a pixel-wise difference for PF and LF runs for the case where there is a particularly high amount of side-branching is shown in Fig.~\ref{fig:absErr}. While the differences are small at any location other than side-branching sites, we do not expect them to ever be — or trend towards — zero at these regions. This is because the evolutionary history of dendritic growth with [physically consistent] thermal noise means that any two runs should never perfectly agree. Indeed, two frames in the pixel-wise comparison such as that in Fig.~\ref{fig:absErr} do not necessarily indicate \textit{statistically invalid} growth patterns; what is important is that both the PF and LF runs are part of the same time-evolution {\it ensemble} corresponding to a given set of process conditions (i.e., $G$ and $V_p$ in this case).

Another important way of quantitatively examining solidification microstructure is observing the power spectra of the interfacial profile and, in the case of directional solidification, the derived Primary Arm Spacing (PAS) \cite{MikePRL,GurevichPRE}. Power spectra of the interface profile offer a better sense of the comparative distribution of the structures at different length-scales within the two systems. Given the stochastic nature of dendritic growth with thermal noise, comparing the power spectra between two “sibling” runs (i.e., runs that share the same initial 65 000 time steps, but that are then run one with PF+AMR and the other with LeapFrog) can be more informative than observing their pixel-wise differences.

To quantify the difference between PF+AMR and LeapFrog power spectra, or Power Spectral Densities (PSDs), and the latter's comparative stability in predictions, we have calculated the $L^2$ norm distance between their respective PSD at each time step. Specifically, we average the mean \& standard deviation over 10 pairs of “sibling” runs. The results are plotted in Figure \ref{fig:L2}. We can observe  significant agreement between both run modalities since the average total $L^2$ norm distance between their PSDs remains, at the very most, ~$10^{-2}$ — mostly staying within the ~$10^{-3}$ order of magnitude.

There is a bulge around $t = 150000dt$ in the left frame of Figure \ref{fig:L2}, which, upon further investigation, is caused by 2 out of the 10 runs in particular where dendrite extinctions occurred at different times in the PF and LF runs (see Supplementary Note 3 for further details). In fact, when these runs are removed from the sample, the $L^2$ norm curves seem to roughly plateau at the steady-state regime ($\sim 150000dt$, see Supplementary Note 3) and then increase sharply around the $200000dt$ mark.

The progressive increase observed in at late time in Fig. \ref{fig:L2} corresponds to where the solidification front — and more importantly, the solute gradient preceding it — encounter the bottom boundary of the system. We hypothesize that the increase in the differences between the power spectra at this point can be attributed to the neural network encountering difficulties learning the specific growth dynamics of the system under finite size effects. These modified growth dynamics would only be observable in the final portions of the training runs, and occupy a comparatively smaller proportion of the training data than the steady state regime, which poses a greater challenge for the neural network to learn.

In Supplementary Note 3, we illustrate the cause of the first hump in the sensitivity of the $L^2$ norm between PSDs as reflecting statistical differences in microstructure that is otherwise  consistent with the same process and material parameters. This metric remains a good approximate indicator of comparative stability, especially in its overall magnitude, but a further examination of its specific contributors are required to fully characterize its fidelity as a metric for assessing microstructure prediction fidelity.

\begin{figure}
    \centering
    \includegraphics[width=0.95\textwidth]{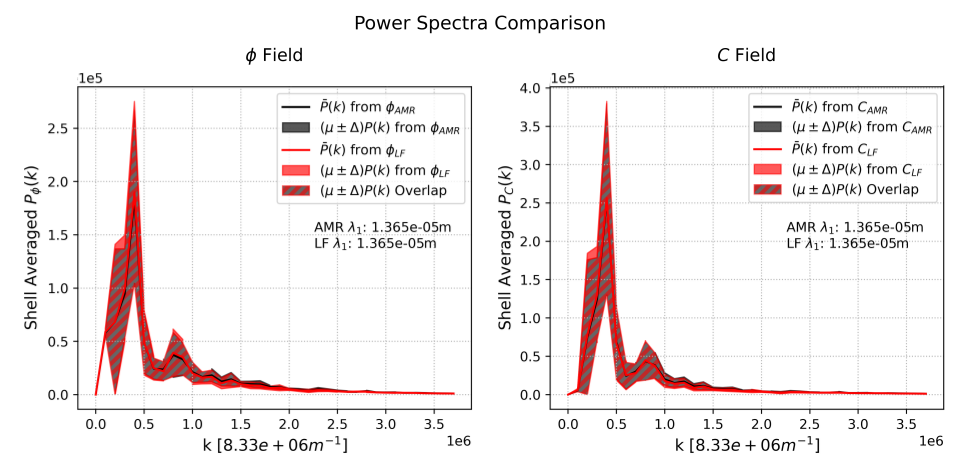}
    \caption{\textit{\textbf{Mean power spectra comparison with uncertainty intervals between direct Phase Field(PF)/AMR  simulations (black) and LeapFrog(LF)-enabled PF simulation (red) for both the phase (left) and concentration (right) fields.} Data is taken at $t=250000dt$. Red/Black barred areas designate regions of overlap between the PF and LF uncertainty. The power spectra have been constructed from averages over 10 simulations with different randomly initialized seeds, and the $k=0$ mode set to 0.}}
    \label{fig:pSpecs}
\end{figure}

As a further illustration of the usefulness of using PSD to characterize microstructure selection in directional solidification, we plot the average power spectra averaged over all sibling runs at their most divergent, i.e. $t=250000dt$, in Fig. \ref{fig:pSpecs} where we observe nearly perfect overlap. The main discrepancy here is at the shoulder around $k=025\cdot8.33\times10^{-6}\textrm{m}^{-1}$.

Observing the differences in  Fig. \ref{fig:absErr} in conjunction with the analysis in Supplementary Note 3, this is most likely associated to the mismatch in dendrite extinction timing occurring around $(x,y) ~ (50,1125)$. It is assumed that they are a result of the different noise patterns brought on by the LeapFrog algorithm and, upon further examination (Supplementary Note 3), do not affect the solidification microstructure as far as its system morphology or length scale selection: There is nearly identical agreement in the ensemble predicted primary arm spacing.

It is noted that primary arm spacing (PAS) is specified through the $k$-mode of the highest peak in the power spectrum densities (PSD). As noted in Fig. \ref{fig:pSpecs}, the PAS obtained from the PF+AMR and LF ensembles are in excellent agreement (an agreement that is present  at any system time — see Supplementary Figure 15). These results further suggest that despite two instances of microstructure, at the same time, not being \textit{identical} as derived from the PF+AMR-evolved system versus the LF system (or even from PF+AMR system alone), the statistical distribution from which system morphology is drawn appears to evolve in a stable and predictable manner. In other words, individual realizations evolve by exploring a statistical ensemble specified by a system process parameters $(G,V)$, system size and initial conditions.


The results of this section further demonstrate the LeapFrog algorithm is capable of predicting the full range of dynamical solidification morphology (i.e, $\phi$ configuration) and solute segregation profiles ($C$ field configuration) in a quantitatively reliable way.

\section{Discussion}\label{sct:discussion}

We have presented a machine learning algorithm that is capable of accelerating multiscale phase field simulations of quantitative directional solidification of two-sided dilute binary alloys simulated with stochastic noise using Adaptive Mesh Refinement (AMR). This is achieved by training a neural network that combines Convolutional Neural Networks, Long-Short Term Memory networks and U-Net architectures to tackle the large system sizes containing numerous time-scales and a large range of length-scales inherent to the microstructure. 

We have also developed a novel loss function for the training of the neural network that dynamically focuses learning on small-scale features of interest by developing cost-effective masks that highlight the locations where said features are present. These masks proportionally amplify small-scale feature contribution in the final loss metric. This approach is particularly beneficial because it allows for the targeting of very small or very faint patterns in the microstructure evolution, with minimal direct calculations based on the fields of the model. This becomes essential to \textit{any} experimentally relevant pattern formation problems where small-scale interfaces are coupled to long-range diffusion to establish structure on larger scales.

We presented an adaptive time stepping algorithm that uses the above neural network (NN) architecture in tandem with a direct phase field (PF) simulator to generate microstructure predictions. These are combined in a “LeapFrog” (LF) fashion, whereby results from the NN are fed into the PF simulator — and vice versa. This LF algorithm is also versatile, since its utilization of ML and PF predictions can be tuned to either favor simulation acceleration or high fidelity with respect to stochastic and fine-grained features, depending on the needs of the application. 

In its current prototypical form, where no explicit effort was made to reduce the overhead of passing information back and forth from the NN to the PF simulator, the combined simulation platform can provide a speed-up rate of at least 5–10 times faster than direct PF simulation. This speed is expected to become much larger when the overhead is factored out. The proposed pipeline could be beneficial when applied to process-microstructure characterization by significantly reducing the total computational time required for the exploration of parameter space — especially if practices such as transfer learning prove applicable.

The speed of the algorithm was also shown to not come at the cost of accuracy, as its predictions — insofar as microstructure statistics are concerned — exactly match those of traditional PF modelling. These include overall morphology, micro-segregation patterns, and primary spacing selection.

Various future directions and extensions of this work are possible. Most directly, the direct integration of the NN into the PF simulator would further increase computational speed up. Furthermore, training the network over a wider range of process conditions (e.g, $(G, V)$) could allow for interpolative predictions within regions of interest in phase space. This would provide the aforementioned speed-ups to simulation generation while \textit{bypassing} the need for initial training data generation (i.e., at new $(G,V)$ points in our application). 

Beyond that, the techniques presented in this work are also not explicitly tied to specific solidification paradigms. The dynamic loss function masking presented in Section ~\ref{sct:LossFctn} can be applied to aid the training of NNs on any system simulations that require interface tracking.

Finally, the network proposed in this work can also be applied to act as a local prediction tool in the so-called “mini-mesh” data structure at the heart of the finite difference scheme of the latest Adaptive Mesh Refinement (AMR) platform\cite{greenwood_quantitative_2018}. This would allow a more direct integration of our NN into AMR, which would further enhance the experimentally relevant system sizes {\it and times scales} accessible to phase field modelling.

\section{Methods}\label{sct:methodology}
\label{methodology}

The section briefly describes the details and phase field (PF) and neural network methodologies, the neural network (NN) algorithm, the NN training modality, and the way the phase field and NN algorithms are combined to form a unified simulation platform whose proof-of-concept is the purpose of this paper.

\subsection{Phase Field Model}\label{sct:PFData}

The PF model generating the data that the neural network will train on and then emulate is that of a  dilute binary alloy simulated through Model C with anisotropy, two-sided diffusion, and an enforced directional thermal gradient $G$ developed by Echebarria et al. (2004)\cite{echebarriaQttvAlloy2004} and pull velocity $V_p$. The reason for this choice of model is its capacity to {\it quantitatively} emulate the conditions and resulting solidification microstructure of many industrial processes such as casting. By exploring and tuning different combinations of $G \textrm{ \& } V_p$, specific patterns, length scales and materials properties can be selected. These microstructural patterns result from evolving the following dynamical equations:
\begin{align}
    \tau A^2(\theta)\frac{\partial \phi}{\partial t} &=  W^2_{\phi}\big [ \nabla \cdot (A^2(\theta)\nabla \phi) \nonumber \\ &-\partial_x[A(\theta)A'(\theta)\partial_y\phi] + \partial_y[A(\theta)A'(\theta)\partial_x\phi]\big ] \nonumber\\ 
    &-H\phi(2\phi-1)(\phi-1) \nonumber \\ 
    &- \lambda \Delta c \left (e^u  + \frac{G(y - V_pt)}{|m_L|\Delta c}-1 \right )g'(\phi) + \eta\label{eq:anisoAndGV}\\
    \frac{\partial c}{\partial t} &= \nabla \cdot \left (D_LQ(\phi)\Delta c\nabla u - \frac{1}{2\sqrt{2}}W_{\phi}e^u \Delta c \frac{\partial \phi}{\partial t}\frac{\nabla \phi}{|\nabla \phi|} \right )\nonumber\\
    &+\nabla \cdot \vec{q} \label{eq:antiTrapping}\\
    u &\equiv \ln \left( \frac{c}{c^l_0[1-(1-k)\phi]}\right) \label{eq:chemPotDiluteFinal}
\end{align}

The main fields being coupled here are the phase field $\phi$ and the concentration field $c$. In the first line of Eq. \ref{eq:anisoAndGV} the characteristic timescale $\tau$ and interface width $W_{\phi}$ determine the scales of the system. In this line and the next we can see the anisotropy $A(\phi)=1 + \epsilon_4\cos{(\theta)}$ which set the preferential growth directions that allow for dendrites to emerge. The third line of Eq. \ref{eq:anisoAndGV} is the double well that sets up the two free energy minima separated by an energy barrier height $H$ that outline the two stable phases (liquid and solid in this case).

The last line of Eq. \ref{eq:anisoAndGV} is the coupling of the phase field, moderated by the coupling constant $\lambda$ and the interpolation function $g'(\phi \rightarrow \{0,1\}) \rightarrow \{0,1\}$, to gradients in the concentration field $\Delta c$. The first term in the brackets in the third line is a measure of the deviation from the equilibrium chemical potential, the second term captures the temperature at a specific height $y$ along the thermal gradient $G$ imposed on the system at a specific time $t$ given the pull velocity $V_p$ and liquidus slope $m_L$. Eq. \ref{eq:antiTrapping} is the time evolution equation of the concentration field where the first term on the right-hand-side modulates solute diffusion scaling by the liquid diffusion/mobility $D_L$, and interpolation function $Q(\phi \rightarrow \{0,1\}) \rightarrow \{1,0\}$, and the gradient and Laplacian in the concentration and reduced chemical potential, respectively. The second term on the right-hand-side of Eq. \ref{eq:antiTrapping} is the solute anti-trapping term introduced by Karma (2001)\cite{karmaPF2001} that counteracts non-physical diffusion effects brought-on by the artificially wide interface width. Finally, the terms $\eta$ and $\nabla \cdot \vec{q}$ at the end of both dynamics equations are added non-conserved and conserved stochastic noise, used in field theoretic models to emulate the role of thermal fluctuations emergent at the atomic scale and which are washed out at the mesoscale.

\begin{figure}
    \centering
    \includegraphics[width=0.95\textwidth]{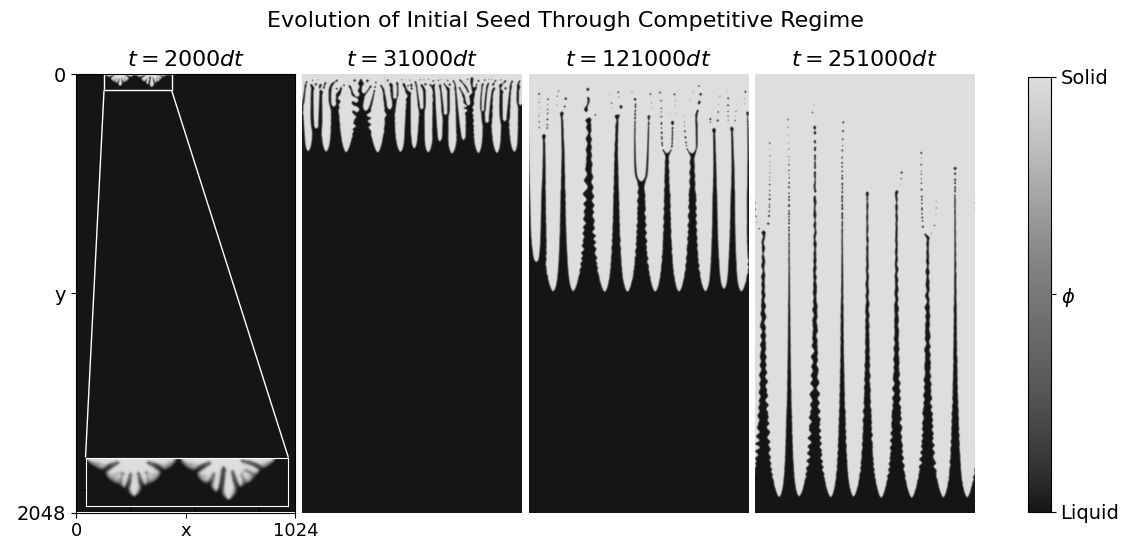}
    \caption{\textit{\textbf{Selection of snapshots from different times in the growth of a dendrite array.}\\ Simulated by Eqs.(\ref{eq:anisoAndGV})~(\ref{eq:chemPotDiluteFinal}) and initiated from the morphology illustrated in $t=2000dt$.}}
    \label{fig:growthRegimes}
\end{figure}

In this work, this above model is simulated on the footprint of a 2048 by 1024 mesh with noise using the Adaptive Mesh Refinement algorithm developed by Greenwood et al. \cite{greenwood_quantitative_2018}. Multiple initial semicircular solid seeds are initialized at the top $y=0$ line and evolved 350 000 time steps, approximately 70 000 time steps past the point where the resulting dendrites reach the $y=2048$ mark, given the chosen pull velocity and time step resolution. These simulations required approximately 3.5 hours to run on 8 AMD Rome 7532 CPU cores at 2.4GHz with 256Mb L3 Caches. 

Given the stochastic nature inherent in solidification microstructure evolution, as well as the fact that said microstructure does not display self-similar scaling in space and time, it would be overly ambitious to frame the task at hand as one of training a neural network algorithm with the above phase field model to predict the final late-stage microstructure directly from process parameters. The relation between process parameters and final microstructure is highly non-linear and would require an extremely large and comprehensive database for a neural network (NN) to learn from.

In our opinion, a more feasible, and practical, goal is to use machine learning to enable the use of larger, and adaptable, simulated time steps. Such an NN would complement traditional adaptive mesh algorithms, leading to dramatic reductions in computational complexity in both the space and time domains. As such, the problem posed to the neural network will be to effectively predict the following:
 
\begin{align}
    \Delta_m \phi(\vec{x}, t_n) \equiv \phi(\vec{x},t_{n+m}) - \phi(\vec{x},t_n) &= \int_{t_n}^{t_{n+m}} \left ( \frac{\partial \phi}{\partial t}\right )dt\label{eq:trainTgtsPhi} \\
    \Delta_m C(\vec{x}, t_n) \equiv C(\vec{x},t_{n+m}) - C(\vec{x},t_n) &= \int_{t_n}^{t_{n+m}} \left ( \frac{\partial C}{\partial t}\right )dt\label{eq:trainTgtsC} \\
    n,m \in \mathbb{Z}^+\nonumber
\end{align}
i.e., advance the evolution of the fields $\phi$ and $C$ from their state at numerical time $t_n$ to their state at $t^{n+m}$, where $m$ corresponds to an integer number of explicit time steps of the original phase field equation. This essentially boils down to emulating Eqs. \ref{eq:anisoAndGV} \& \ref{eq:antiTrapping} over $m$ time steps in one [adaptable] time step.

As a proof of concept of the above idea, this work focuses on the competitive cellular and dendritic growth regime of these systems and ignore the transient regime wherein the solidification front transitions from its initial condition to dendrite fingers. This regime sets the foundation for the complex final solidification microstructure that forms in practical alloys, as well as numerous other first order phase transformations driven by competitive cellular arrays. To make this criterion quantitative, phase field simulation data over the time period $t \in [65000,250000]dt$ is be used as samples for the training regiments of the machine learning algorithm proposed in this work.

\subsection{Neural Network Architecture}  \label{sct:MLArch}

The choice of neural network architecture(s) as well as the loss function that they train on depend greatly on the problem tackled. In this section, we describe how the multiscale nature of dendritic solidification informed our choices on the former. 

Requiring a machine learning model to predict the interaction of dendrite cells, the emergence of entire side-branches or dendrite extinctions, etc., from input snapshots tens of thousands of time steps before such events even begin — let alone directly predicting final microstructure — poses an enormous challenge. It would require the neural network model to directly predict emergent behaviour in a highly non-linear system; moreover, unlike a second order transition, microstructure arising in a first order transition such as solidification cannot be renormalized under a simple set of scaling laws. 

A key challenge in developing any neural network predictor of solidification microstructure evolution over some time frame thus becomes how to correlate the outputs of Eqs.~(\ref{eq:anisoAndGV})-(\ref{eq:chemPotDiluteFinal}) to compound errors made by the Neural Network (NN), such that the output from the NN remains, at any time, within a statistical {\it ensemble} of solutions consistent with the predictions of the original phase field equations.  

Given the highly non-linear and stochastic nature of dendritic evolution, we have developed our NN to act as an aid to the PF time-evolution code, a more feasible and reliable objective. This PF aid function consists of predicting the changes in the system's fields from an input time $t_n = n\cdot dt$ to a projected time $t_{n+m} = (n+m)\cdot dt$, where $dt$ is the physical time interval and $n,m \in \mathbb{Z}^{+}$ integers. As a concrete example, the value $m=1000$ was chosen here as over that time span microstructure in the system has time to evolve meaningfully without at the same time spawning entirely new patterns or behaviours.

We employ Long-Short Term Memory (LSTM) networks, a flavor of Recurrent Neural Networks (RNNs) presented by Hochreiter \& Schmidhuber 1997\cite{hochreiter_long_1997}, which are directly suited to the task of making time-evolution predictions taking into account multiple time-scales. Given a time-series of a system, its Long Term Memory (LTM) and Short Term Memory (STM) pipelines allow it to separately and respectively track consistent trends in the data (e.g., front advancement, coarsening) as well as more immediate ones (e.g., solute diffusion, side-branch growth). This network architecture has been applied in the phase field domain to spinodal decomposition (de Oca Zapiain, Stewart \& Dingreville 2021\cite{montes_de_oca_zapiain_accelerating_2021}, Hu, Martin \& Dingreville 2022\cite{hu_accelerating_2022}), epitaxial growth (Qin et al. 2023, in combination with an attention layer\cite{qin_grainnn_2023}), and simple, single-field descriptions of patters emerging in 1st order phase transitions (Peivaste et al. 2022\cite{peivaste_machine-learning-based_2022}).

LSTMs require additional modification, however, as they were initially designed for the parsing of low-dimensional vectorized data. It is thus necessary to introduce downsampling and upsampling Convolutional Neural Networks (CNNs) (LeCun, 1998\cite{lecun_gradient-based_1998}) before and after the LSTM (respectively) as well as a convolutional kernel within it. This integration leverages the translationally invariant learning of CNNs as well as their reduced relative memory requirements to allow the network to process the fairly large systems sizes we tackle in phase field modelling. The result is termed a Convolutional Long-Short Term Memory (CLSTM) network.

Finally, the last ML architecture we incorporate is the U-Net architecture (Ronnenberg, Fischer \& Brox 2015\cite{ronneberger_u-net_2015}). CNNs by themselves were not found to be able to keep track of all the length-scales involved in dendritic solidification. U-Nets are particularly well-equipped for this multiscale context. These networks divide the dataflow within the NN into separate downsampling (encoder) and upsampling (decoder) stages that take the input data down to some smallest resolution and back to its input resolution. By itself, this does not differ from the downsampling/upsampling pipeline achievable with CNNs. However, U-Nets have “skip connections” between encoders and decoders that operate at similar resolutions, which allow for the transmission of fine-grain pattern information that would usually get lost during compression to lower resolutions. 

Versions of this network structure have been recently used in combination with various “core” architectures by Dingreville et al. 2023 \cite{oommen_rethinking_2023} to evolve systems of physical vapor deposition. Notably, however, an LSTM core is not explored in their implementation. This is a significant incorporation that we found is necessary to accurately identify and emulate the multiple time scales of dendritic solidification. The LSTM core conveys additional capabilities to the overall network and resulting simulation algorithm, which will be discussed in Section \ref{sct:LF}.

\begin{figure}[t]
    \centering
    \includegraphics[width=0.95\textwidth]{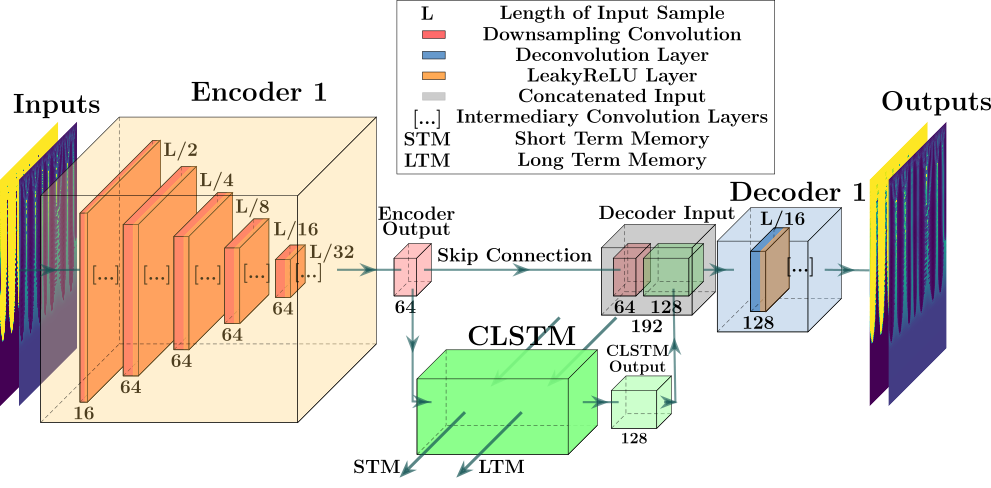}
    \caption{\textit{\textbf{Network diagram of Multiscale Convolutional Long-Short Term Memory (MSCLSTM) network with only one Encoding/Decoding stage.} One by one and in order, the inputs from a time-sequence are encoded, evolved by the CLSTM core, and decoded. The CLSTM generates the Short Term Memory and Long Term Memory states for the next time-step. A skip connection between the encoder and decoder in this arrangement transmits a fairly compressed data stream.}}
    \label{fig:1stStage}
\end{figure}

A basic instantiation of the network that illustrates the general structure can be seen in Fig. \ref{fig:1stStage}, which shows the ingestion of one frame of the system's evolution at $t_n$ that gets processed and compressed by the Encoder block, evolved through the CLSTM, and finally decompressed by the Decoder block to output $\Delta_m \{\phi, C\}(\mathbf{x},t_n)$ that will evolve the system fields to $t_{n+m}$. During the evolution by the CLSTM, the Short Term Memory (STM) and Long Term Memory (LTM) states are generated that will inform the evolution of the next snapshot of the system. This sequence is repeated for every snapshot of the input data.

\begin{figure}[t]
    \centering
    \includegraphics[width=0.95\textwidth]{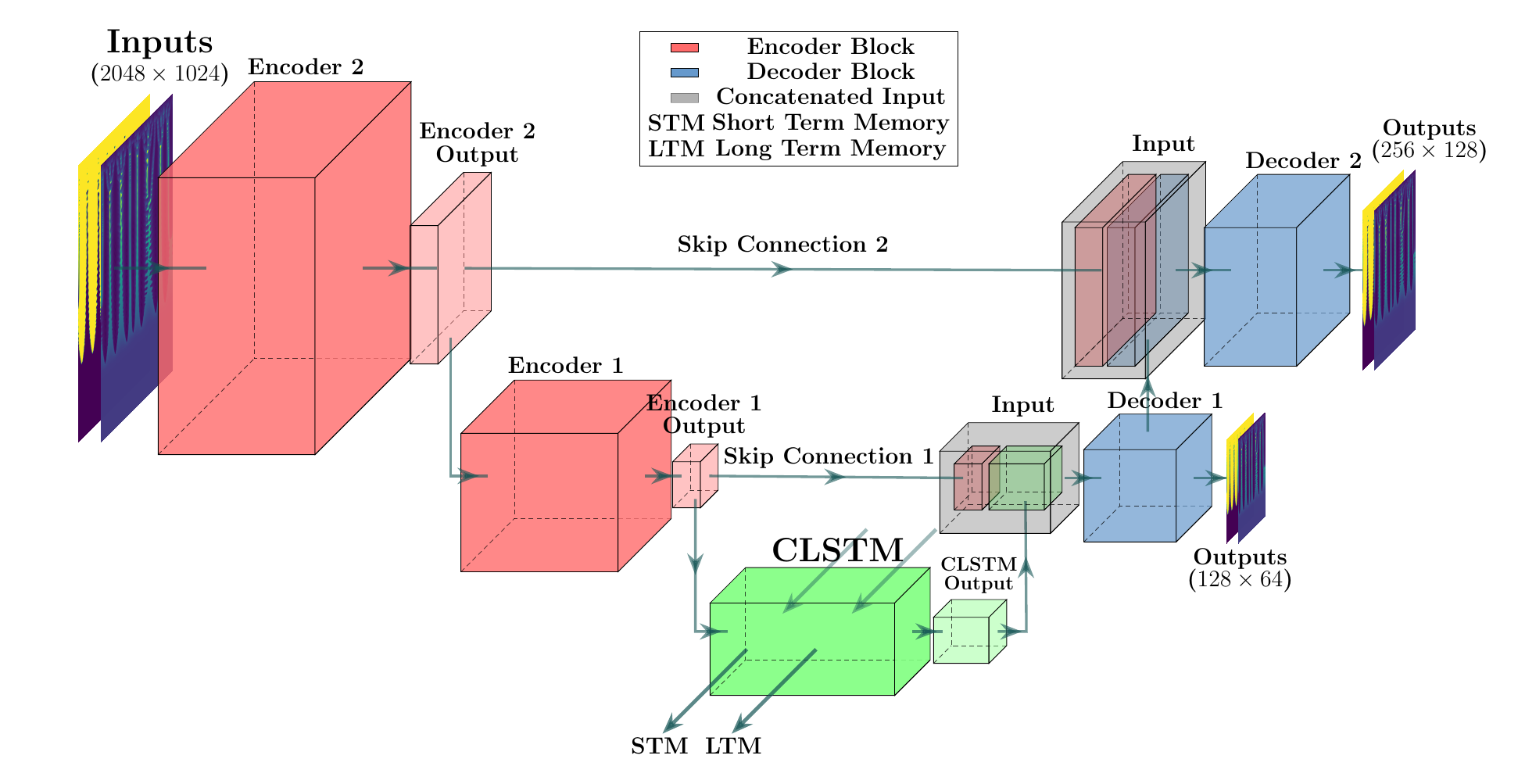}
    \caption{\textit{\textbf{Network diagram of an MSCLSTM with two Encoder/Decoder stages.} The data flow is the same as in Fig. \ref{fig:1stStage}, however now Skip Connection 2 offers a data stream through which signals can pass without being subjected to the highest degree of compression in the network. Notably, every Decoder outputs a system prediction at its own resolution that can be compared to coarsened targets in order to generate loss signals tailored to each resolution/length-scale.}}
    \label{fig:2ndStage}
\end{figure}

The network in Fig. \ref{fig:2ndStage} is an expansion of the network in Fig. \ref{fig:1stStage}, with the addition of one U-net stage and skip connection, which we use to allow process of larger data inputs. The higher skip connection (“Skip Connection 2”) transmits data to the second Decoder before the dimensionality reduction carried out by “Encoder 1”, and thus contains higher resolution system information. 


The final trained version of the network described in this subsection is composed of 5 U-net and skip connection stages. The NN's final number of U-Net \& skip connection stages is a result of the lack of improvement in the results of training sessions including additional layers, as well as the increased computational cost involved to train a larger network.

Said training sessions were performed following a grid search approach for hyperparameter exploration. This method was made feasible given the high number of constraints imposed on the depth of the network by the longest length-scale present in the problem being tackled (See also Ref.~\cite{Pinto_MSC}), as well as those imposed by the interplay between data characteristics and hardware limitations.

Both the number of layers and the number of snapshots included in an input sample were correlated with significant increases in the memory requirements of the weight gradients to be computed during backpropagation.  Due to memory constraints imposed by GPU capacity (NVIDIA A100 with 40Gb of VRAM), the MSCLSTM in our application is trained on 5 snapshot inputs, i.e., $\mathbf{I}_n \equiv [t_{n-4m}, t_{n-3m},...,t_{n}]$. Despite the quite respectable VRAM capacity of the NVIDIA cards, as well as the significant reduction in memory costs allowed by PyTorch's Adaptive Mixed Precision package \cite{micikevicius_mixed_2018}, managing the space required to store the network, the samples, and especially the gradients from backpropagation remains a significant challenge. 

The training samples $\mathbf{I}_n$ are sampled within the time range $t \in [65k,250k]\cdot dt$ described at the end of Section \ref{sct:PFData} from 11 runs simulated through traditional PF+AMR software, with 3 additional runs used for testing and 1 run for validation. In total, the network requires approximately 7 hours to train (although, this training time allows a generous tail to the loss curve that could very feasibly be reduced). Each simulation used here required 3.3 hours to produce, however this time will not be included in the time comparison between the AMR+PF pipeline and the LeapFrog pipeline. This is because the data produced as part of the training data can double up as data used for the final microstructure characterization of a parameter space exploration.

With this iterative process, we can construct a prediction for the system at $t_{n+m}$ which can then \textit{also} be used as a new input to produce $\Delta_m\{\phi,C\}(\mathbf{x},t_{n+m})$, which can itself be used to produce $\{\phi,C\}(\mathbf{x},t_{n+2m})$ , and so on. Again, due to memory constraints, we use a training algorithm that produces the output set $\mathbf{O}_n$ necessary to make three $1000dt$ time span jumps (i.e.,  $\mathbf{O}_n \equiv \Delta_m\{\phi,C\}[(\mathbf{x},t_{n-4m}),...,(\mathbf{x},t_n),(\mathbf{x},t_{n+m}),(\mathbf{x},t_{n+2m}),(\mathbf{x},t_{n+3m})]$). Larger memory resources would allow larger time span predictions of the field configurations.

\subsection{Loss Functions} \label{sct:LossFctn}

\begin{figure}
    \centering
    \includegraphics[width=0.95\textwidth]{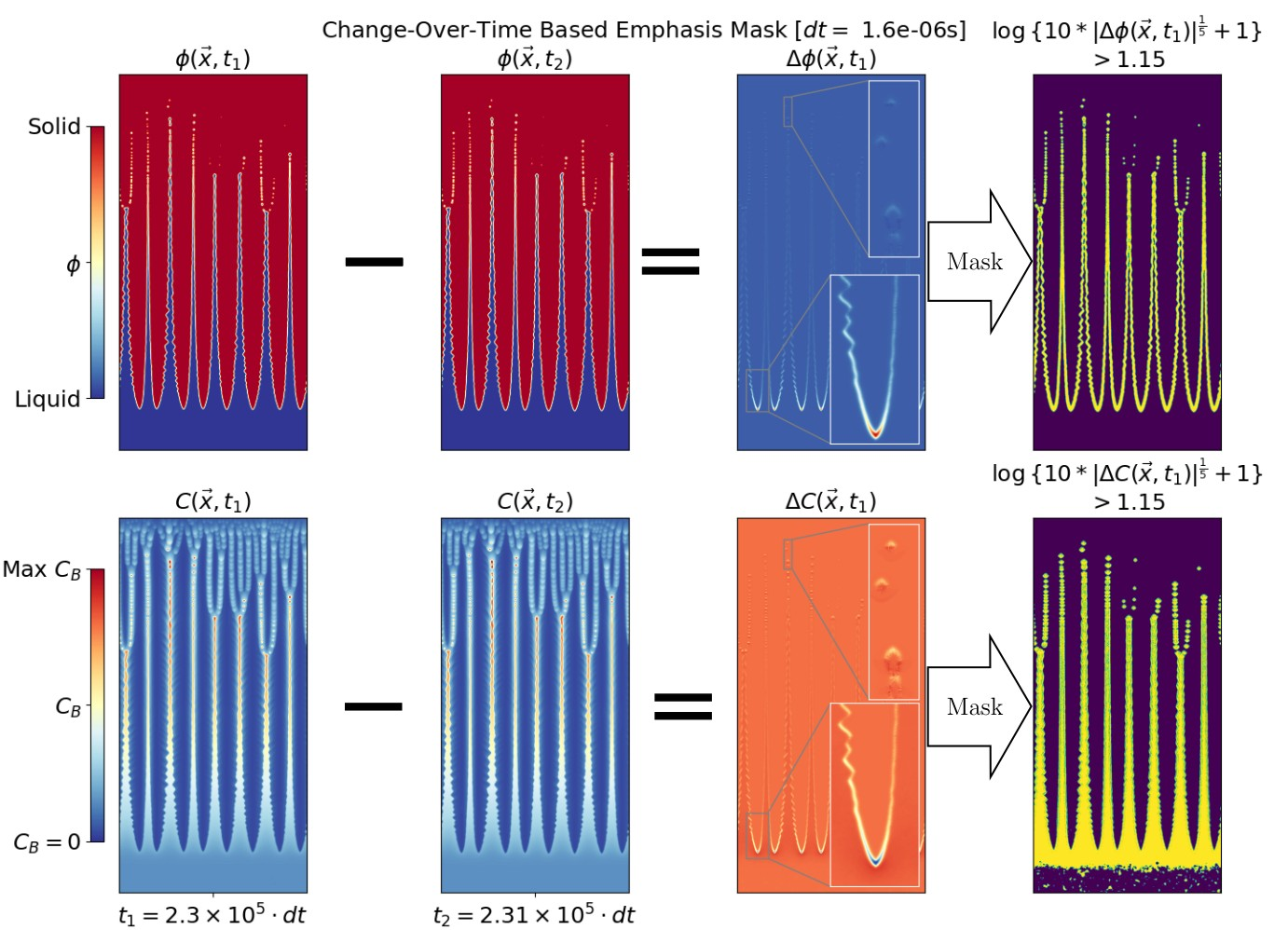}
    \caption{\textit{\textbf{Interface \& Gradient mask generation.} Taking the difference between the system's fields at two different times (first two columns) to construct the left-hand sides of Eq. \ref{eq:trainTgtsPhi} \& \ref{eq:trainTgtsC} (third column). It is clear that values of 0 occupy the majority of the $\Delta \{\phi, C\} (\mathbf{x},t)$ fields (blue in the case of $\phi$, and orange in the case of $C$). The final column results from the construction of a binary mask after the application of a custom re-scaling function to emphasize the interface and solute diffusion.}}
    \label{fig:dtMask}
\end{figure}

A simple training algorithm that trains the network with the schema described above using the ubiquitous Mean Squared Error (MSE) loss function falls into some “local minima” in the solution space of network configurations. That is to say, the network will train to emulate the “lowest hanging fruit” to quickly minimize its error metric — in this case: the bulk phases. The network predicting $t = 121000 dt$ in Fig. \ref{fig:growthRegimes} will quickly learn to paint the top half as entirely solid and the bottom half as entirely liquid, ignoring the comparatively minuscule complex interfacial patterns — no matter their importance in PF. 

Typically, in ML, the number of training samples containing difficult-to-learn features is increased in the training set such that the conceptual importance of said features is adequately represented. In the case of solidification, however, \textit{all} samples contain a solidification interface that is orders of magnitude smaller than the microstructurally important features whose dynamics it governs through a coupling with the diffusion field, which typically spans longer length scales. Thus, in large systems, the interface comprises an important system volume too small to properly incentivize the network to pay attention to it during training.

To address this, we develop a dynamic correction for the asymmetry in feature focus through a “feature weighted loss function” $\mathcal{L}_{fw} (\mathbf{x},t)$. The main components of this process are to first identify features that are neglected, construct a binary map $\mathbf{M}$ of their presence in the system, and finally to multiply the MSE loss contribution from those pixels by a factor $\alpha'$, which yields, 
\begin{align}
    \mathcal{L}_{fw}(\mathbf{x}, t) &\equiv \Bar{\mathbf{M}}\cdot \mathcal{L}(\mathbf{x},t) + \alpha' \mathbf{M} \cdot \mathcal{L}(\mathbf{x},t)\label{fwLoss} \\
    \alpha' &\equiv \frac{\alpha N}{\sum\mathbf{M} + \epsilon}, \epsilon = 1\times10^{-6},
    \label{alpha}
\end{align}
where $\mathcal{L}(\mathbf{x},t)$ is the original loss function(s) chosen, $\Bar{\mathbf{M}}$ is the constructed feature mask passed through the logical “NOT” operator, $N$ is the total number of pixels in the entire sample and $\sum\mathbf{M}$, which denotes the sum of the elements of (the array) $\mathbf{M}$, returns the number of pixels that the feature being targeted occupies within the sample. This form of $\alpha'$ adaptively scales the numerical emphasis based on the relative system occupation of the feature, while avoiding divisions by 0 if it is absent. The numerical factor $\alpha$ is an \textit{ad-hoc} tunable factor. In our case, it has a value of 100 given the very faint numerical presence of the solute diffusion in the concentration field highlighted by the mask generated in the final column of Fig. \ref{fig:dtMask}.

The benefits of this method are that masks that target relevant features can be constructed with simplicity and versatility, but more complex mask constructions can also be implemented in a manner that keeps the computations necessary for their construction outside the gradient tree.

In addition to the mask illustrated in Fig. \ref{fig:dtMask} being applied to the MSE loss, a SoftDice loss term\cite{milletari_v-net_2016} is added as well as another feature emphasis is on the concentration profile evolution in areas where the target fields have an order parameter value $\phi > 0.5$ is incorporated into the training (see Equations 4-8 in Supplementary Note 2). The former has been found to facilitate image segmentation  in U-Nets used for medical imaging in the presence of significant noise, and the latter corrected for an observed compounding error in the predicted concentration profile of the dendrites, while also coincidentally incentivizing mass conservation. Previously, an explicit penalty for not respecting mass conservation across the system was included, but was later removed since it became redundant.

A network trained on the feature weighted loss function presented above results in a NN capable of generating high-fidelity predictions. Only some small numerical oscillations in the concentration field remain, which can be easily corrected for using post-processing techniques described in Section \ref{sct:predQlty}.

In closing this section, it is noted that an attempt was made to inform the network on the dynamics and physics of the system — such as including a power spectra comparison term in the loss. The latter was found to improve the length-scale correspondence between the PF+AMR predictions and the LF predictions (see Supplementary Note 2) by a marginal amount, while increasing the GPU memory utilisation by approximately 4Gb.

The memory usage increase is mainly attributed to the large number of operations per pixel that such physical losses can imply, and Supplementay Note 2 illustrates how using a convenient intermediary to the calculation of the full physical metric can provide similar benefits while eschewing the full calculation of said metric (in this case, simply using the magnitude squared of the FFT of the network outputs).

Given the memory constraints mentioned at the end of Section ~\ref{sct:MLArch}, the aforementioned MSE, Soft Dice and solid concentration were kept over the power-spectrum-based loss due to their more sizable impact to the overall training and final implemented algorithm results.



\subsection{Adaptive LeapFrog Algorithm}\label{sct:LF}

The main algorithm developed for this work is designed to couple the MSCLSTM algorithm presented above with direct simulation of Eqs.~(\ref{eq:anisoAndGV})-(\ref{eq:chemPotDiluteFinal}) to increase the efficiency of simulating microstructure evolution. Specifically, direct PF simulation is used to compute output data from the initial seed up to the early stages of the dendritic regime, after which this output data is input into the trained MSCLSTM. After a set number of time very rapid jumps by the NN amounting to $N_{ML} = n_{ML}\cdot m\cdot dt$ time steps (or, possibly, after a monitored error metric surpasses a threshold of acceptability), the result of the neural network's time jumps is handed back to the PF code, which can correct any errors accumulated during the MSCLSTM prediction phase, as well as reintroduce explicit thermal fluctuations compared to the  $N_{PF}$ time steps, which were evolved without this component of the phase field dynamics. Following this, the NN then takes over again and the above cycle repeats until the solidification front reaches the end of the simulation system (i.e., sample size). The two time-evolution methods each jump in from where the other left off in what we term a “Leapfrog” manner.

It is noted that the implementation of the hand-off between the PF and ML described above is presented in this work as a proof-of-concept, and as such several key features remain rudimentary and are being  improved for future work. Some of these are discussed next. 

On the scripting front, it is currently simply a “for” loop in a bash script that call the PF simulator, run in C++, and then the ML predictor, run in Python. 

On the logging front, as signaled in Section \ref{sct:PFData}, the direct PF simulator here is being implemented using Adaptive Mesh Refinement (AMR). The network, however, is trained on systems represented on a uniform mesh. As a result, once the PF simulator is done its pre-determined simulation period, it must map its output onto a uniform mesh representation. Conversely, when the ML predictor is done making its predictions, it must hand its output back to the AMR algorithm at the \textit{highest} level of mesh refinement, after which the AMR algorithm  spends the first few time steps coarsening the mesh where appropriate.

A final feature that can be made more efficient is the practice of regular time-stepping. System snapshots are presently generated in numbers whose interval is sufficient to generate an input for the ML algorithm matching the format of the training samples. This is not \textit{strictly} necessary, but including more snapshots in the input will be likely be of no benefit, as the network has not trained to track trends on those time scales. It would be possible to use an input sample that is shorter than the training samples, however this would run a risk of not providing enough data on the temporal timescale required for the network to construct a complete LTM state.

Despite these considerations, the LeapFrog scheme remains a versatile and efficient algorithm thanks to its LSTM core. Since LSTMs always produce internal Short Term Memory (STM) and Long Term Memory (LTM) states, we do not have to adhere to the exact number of predictions it was tasked with making during training. The ratio of $N_{ML}\textrm{ to }N_{PF}$ can thus be adjusted either to leverage time-savings (increase $N_{ML}$), or simulation fidelity (increase $N_{PF}$). In this way, the LeapFrog algorithm can act as an \textit{adaptive} time-stepping acceleration method that allows for significant time-savings.

\section*{Data Availability}

The datasets used and/or analyzed during the current study available from the corresponding author (damien.pinto@mail.mcgill.ca) upon reasonable request.

\section*{Code Availability}

The underlying code for this study and training/validation datasets are not publicly available, but may be made available to qualified researchers on reasonable request from the corresponding author (damien.pinto@mail.mcgill.ca).

\section*{Acknowledgements}

We acknowledge support and funding from the Mitacs Organization through its Accelerate Fellowship program and IREQ-Hydro-Québec. We also acknowledge funding from the Canada Research Chairs program, Grant ID 235492 .  Finally, we acknowledge the computational resources made available by the Digital Research Alliance of Canada.

\section*{Author Contributions}

D.P. designed and trained the ML model, generated the data for model training and validation, implemented the LeapFrog algorithm, performed analyses of the results and wrote the paper. M.G. provided PF AMR code, found simulation parameters that demonstrated significant side-branching. M.G. and N.P advised the ML model and LeapFrog algorithm design. N.P. supervised all work, and co-wrote the paper.

\section*{Competing Interests}

The authors declare no competing financial or non-financial interests.

\section*{Additional Information}

\textbf{Correspondence} and requests for materials should be addressed to Damien Pinto or Nikolas Provatas.

\noindent \textbf{Reprints and permission information} is available at \href{https://www.nature.com/nature-portfolio/reprints-and-permissions}{nature.com/nature-portfolio/reprints-and-permissions}.

\bibliographystyle{ieeetr}
\bibliography{bibliography}
\ifarXiv
    \foreach \x in {1,...,\numbersupplementpages}
    {
        \includepdf[pages={\x}]{\supplementfilename}
    }
\fi

\end{document}